\documentclass[manuscript,screen]{acmart}
\usepackage{enumitem}
\usepackage[sorting=none]{biblatex}
\usepackage{xcolor}
\usepackage{lipsum}

\newcommand\blfootnote[1]{%
  \begingroup
  \begin{NoHyper}
  \renewcommand\thefootnote{}\footnote{#1}%
  \addtocounter{footnote}{-1}%
  \end{NoHyper}
  \endgroup
}

\usepackage{hyperref}
\hypersetup{
    pdftitle={Infrastructure-Native Computing with Electric Power Grids},    
    pdfauthor={},    
    pdfsubject={},   
    pdfcreator={},   
    pdfproducer={},  
    pdfkeywords={},  
    colorlinks=false,         
    linkcolor=blue,           
}



\begin{document}

\title{Infrastructure-Native Computing with Electric Power Grids}

\author{Yubo Song}
\email{yuboso@energy.aau.dk}
\orcid{0000-0003-1452-449X}
\author{Subham Sahoo}
\email{sssa@energy.aau.dk}
\orcid{0000-0002-7916-028X}
\author{Freja Basse}
\email{ffhk@energy.aau.dk}
\orcid{0009-0003-0293-4173}
\affiliation{%
  \institution{Department of Energy, Aalborg University}
  \city{Aalborg}
  \country{Denmark}
}

\blfootnote{
e-mails: \texttt{\{yuboso, sssa, ffhk\}}@energy.aau.dk
}

\begin{abstract}
Computing is conventionally implemented by hardware engineered for information processing. Here we investigate infrastructure-native computing: the use of a physical system built for another primary function as a fixed computational operator. In time-domain simulations of an IEEE 14-bus electrical network, Kirchhoff’s current law and Ohm’s law relate voltage-reference perturbations applied at distributed controllable nodes interfaced by power electronics converters to current responses through a topology-dependent transformation. A trained digital encoder and decoder exploit this transformation for image classification, reaching 91.5\% accuracy on MNIST and 82.25\% on Fashion-MNIST. The modeled operator is represented by 933 surrogate parameters, compared with 12,340 task-trained parameters for an accuracy-matched fully connected core transformation. Current superposition further supports concurrent spatial sharing of the operator and sequential temporal reuse, with per-stream accuracies above 85\% and 93\%, respectively, in surrogate-model evaluations. Evaluations on CIFAR-10 and repeated 10-class tasks sampled from a Butterflies-and-Moths dataset show that the incremental utility of the physical operator depends on the representation supplied by upstream digital feature extraction. These results provide a simulation-based proof of concept for infrastructure-native computing with electrical networks and identify topology, accessible control channels, and input representation as determinants of its computational utility.
\end{abstract}

\renewcommand\keywordsname{Key Words}
\keywords{Physical Computing | Electric Power Grids | Artificial Intelligence | Smart Energy Systems | Sustainable Infrastructures}

\maketitle


\section*{Introduction}

The rapid growth of artificial intelligence is increasing the computational demand for efficient information processing, motivating computing paradigms that exploit physical dynamics rather than conventional digital arithmetic. Physical neural networks have demonstrated that optical \cite{xue2024fully,lin2018all}, acoustic \cite{momeni2023}, and microwave substrates \cite{del2018leveraging,sol2024experimentally,abou2025programmable} can perform useful transformations directly through their native physical dynamics. These approaches typically rely on physical substrates that are deliberately engineered or selected for computation, with their structure, parameters, and operating conditions designed to support information processing \cite{chen2025optical,a2025self}. An alternative is to exploit computational structure already present in physical infrastructure developed for other purposes. Large-scale engineered systems continuously evolve under governing physical laws, imposing structured relationships between distributed states and responses that may be computationally useful without being explicitly designed for that function \cite{yan2024emerging,nakamura2026ensemble}.

Electric power grids provide a physical substrate in which computational transformations are already embedded in infrastructure designed for energy delivery. They are large-scale, continuously operating networks whose states are governed by conservation laws and accessed through a sparse set of highly controllable nodes as enabled by power electronics converters. Modern grid operation already supports controlled voltage perturbations through converter-interfaced resources, providing an operational pathway for injecting small signals without modifying the underlying network topology \cite{entsoe2024report,entsoe2025report}. We exploit this controllability to encode inputs as voltage perturbations and read out the resulting current responses, as illustrated schematically in Fig.~\ref{fig_1}a and for the IEEE 14-bus topology in Fig.~\ref{fig_1}b \cite{IEEE14BusData}. Under steady-state operation, Kirchhoff's current law (KCL) and Ohm's law give
\begin{equation}
    \Delta \mathbf{I} = \mathbf{Y} \Delta \mathbf{V},
\end{equation}
where $\mathbf{Y}$ is the network admittance matrix. The grid therefore implements a fixed linear transformation between distributed inputs and outputs, with $\mathbf{Y}$ determined by network topology and line impedances rather than learned from task data.

This physical transformation provides a structured computational operator before any task-specific learning takes place. The Jacobian $\partial \mathbf{I}_{\mathrm{rms}}/\partial \mathbf{V}_{\mathrm{ref}}$ is determined by the network geometry and impedance, producing a sparse pattern of coupling strengths between controllable nodes and read-out locations \cite{song2025unlocking, sahoo2025spike}. Electrically proximate nodes therefore interact through stronger physical couplings than electrically distant nodes, embedding prior structure into the transformation before classification training begins. In this sense, the electric grid provides an \textit{inductive bias} through its physical topology: relationships that a fully connected digital transformation must infer from data are constrained \textit{a priori} by the network geometry. In the IEEE 14-bus system, the physical operator achieves comparable classification performance to an accuracy-matched fully connected transformation while requiring 933 represented parameters compared with 12,340 for the corresponding digital transformation, corresponding to a 13.2-fold reduction in operator representation \cite{han2015learning, sharma2026physics}. The resulting topology-dependent physical operator is visualized through the Jacobian structure in Fig.~\ref{fig_2}a.

The linearity of the physical operator provides a second computational property: \textit{multiplexing of multiple inputs on shared infrastructure}. For independently encoded perturbations,
\begin{equation}
    \mathbf{Y}(\Delta \mathbf{V}_1+\Delta \mathbf{V}_2)
    =
    \mathbf{Y}\Delta \mathbf{V}_1 + \mathbf{Y}\Delta \mathbf{V}_2,
\end{equation}
so their contributions remain additive at the read-out. We exploit this property in two modes. In \textit{spatial multiplexing}, different inputs are assigned to different controllable power electronic nodes and read-out channels; whereas in \textit{temporal multiplexing}, successive inputs occupy distinct encoding windows and are decoded separately. Spatial partitioning reduces the effective set of physical couplings available to each task, whereas temporal multiplexing allows successive tasks to access the full network operator. Multiple inference streams can therefore share the same physical computational layer without replicating its physical operator. The encoding, propagation, and read-out processes underlying this architecture are illustrated in Fig.~\ref{fig_1}a and Fig.~\ref{fig_1}c.

We test the physical operator in the IEEE 14-bus system using a lightweight finite-impulse-response (FIR) encoder and a linear decoder. The resulting hybrid system achieves 91.5\% accuracy on MNIST \cite{mnist} and 82.25\% on Fashion-MNIST \cite{fashion-mnist}, while the physical operator itself contains no task-trained parameters. Its input--output transformation is fixed by the grid topology and can be directly characterized from the response to controlled voltage perturbations. The distinction between the physical operator and its digital surrogate is further resolved through robustness experiments: classification remains stable under the tested line-impedance and load variations and under measurement noise up to $0.05$~pu, whereas change in network topology can strongly degrade performance. The difference between the ideal surrogate and the time-domain simulation is examined through controlled perturbations of network parameters and readout noise. The resulting sensitivity analysis identifies perturbation levels that produce accuracy changes comparable to the observed surrogate–simulation gap, while not uniquely attributing that gap to a single source. Representative physical responses and classification outputs are shown in Fig.~\ref{fig_1}c and Fig.~\ref{fig_1}d.

We then examine how the contribution of the physical operator depends on the representation supplied to it. We evaluate CIFAR-10 \cite{cifar10} and repeated ten-class classification tasks sampled from a 100-species Butterflies-and-Moths \cite{butterfly}, comparing the physical-operator pathway with a static linear-pass control that receives the same encoded features. Across these evaluations, the incremental contribution of the physical transformation is larger when the encoded representation is less readily separated by the static control and becomes smaller when more expressive upstream feature extraction or pre-processing makes the representation easier to classify. These experiments do not test the grid as a stand-alone visual feature extractor; rather, they assess the role of a fixed physical operator within hybrid digital–physical pipelines.

Together, these results show that useful computation can be extracted from physical infrastructure without redesigning the infrastructure as a computer: the network topology determines the computational operator, conservation laws provide its native superposition, and the structure of the encoded data determines when that operator contributes useful computation.

\section*{Results}

\subsection*{Electric Grids Implement a Fixed Topology-Dependent Physical Operator}

We first establish the physical input-output transformation that serves as the computational substrate. Around a nominal operating point, voltage-reference perturbations applied through converter-interfaced nodes produce current responses governed by the network topology, line impedances, converter and network dynamics. In steady-state, this relationship reduces to the nodal admittance mapping introduced above:
\begin{eqnarray}
    \Delta \mathbf{I} = \mathbf{Y} \Delta \mathbf{V},
    \label{eq_ohm}
\end{eqnarray}
where $\mathbf{Y}$ is the nodal admittance matrix. In time-domain, the effective transformation from $\Delta \mathbf{V}_\mathrm{ref} (\mathrm{t})$ to $\Delta \mathbf{I_{rms}} (\mathrm{t})$ additionally includes the modeled distributed energy resource (DER), interfacing converter and network dynamics. In either case, the physical transformation is fixed by the electrical system and is not optimized using classification labels. We assess the contribution of the physical operator by comparing the hybrid pipeline with a static-pass control that receives the same encoded input but bypasses the network transformation. This comparison tests whether the fitted electrical-network response provides predictive information.

To use this physical transformation for inference, we construct the hybrid physical-digital architecture shown in Fig.~\ref{fig_1}a. A lightweight finite impulse response (FIR) encoder \cite{oppenheim1999discrete} maps each input image to time-varying voltage perturbation waveforms $\Delta \mathbf{V}_\mathrm{ref} (\mathrm{t})$, which are applied to four DER nodes 1, 2, 3 \& 6 (marked in blue in Fig.~\ref{fig_1}a) in the IEEE 14-bus system. The FIR encoder is deliberately kept lightweight to isolate the contribution of the physical operator from that of a deep learning-based front end. Its linear filtering structure also preserves a largely linear signal path between the input representation and the voltage perturbations applied to the grid. The physical grid layer then processes these perturbations through its native power flow dynamics, governed by Eq.~\ref{eq_ohm}, and the resulting current responses $\Delta \mathbf{I}_\mathrm{rms} (\mathrm{t})$ are sampled at the read-out DER node 8 (marked in red in Fig.~\ref{fig_1}a), and a linear decoder maps the resulting time-stamped response vector to class probabilities $\mathbf{P}\{\texttt{label}\}$ in Figs.~\ref{fig_1}a,c.

The physical operator itself contains no task-trained parameters. Its input-output transformation is determined by the grid topology and impedances, and can be characterized directly from the response to controlled voltage perturbations. During offline training, a differentiable, geometry-aware surrogate of the IEEE 14-bus system is used to reproduce this physical transformation and update the encoder parameters. The surrogate operator is fitted to reproduce the simulated input–output response and is then held fixed while the encoder and decoder are trained for classification. Its 933 parameters are therefore surrogate-identification parameters, not task-trained classification weights. The complete hybrid classifier nevertheless contains additional task-trained digital encoder and decoder parameters, reported in Table~\ref{tab_par}. This separation allows the contribution of the physical transformation to be assessed independently of the learned encoder and decoder.

\clearpage
\begin{figure}[p]
    \centering
    \includegraphics[width=\linewidth]{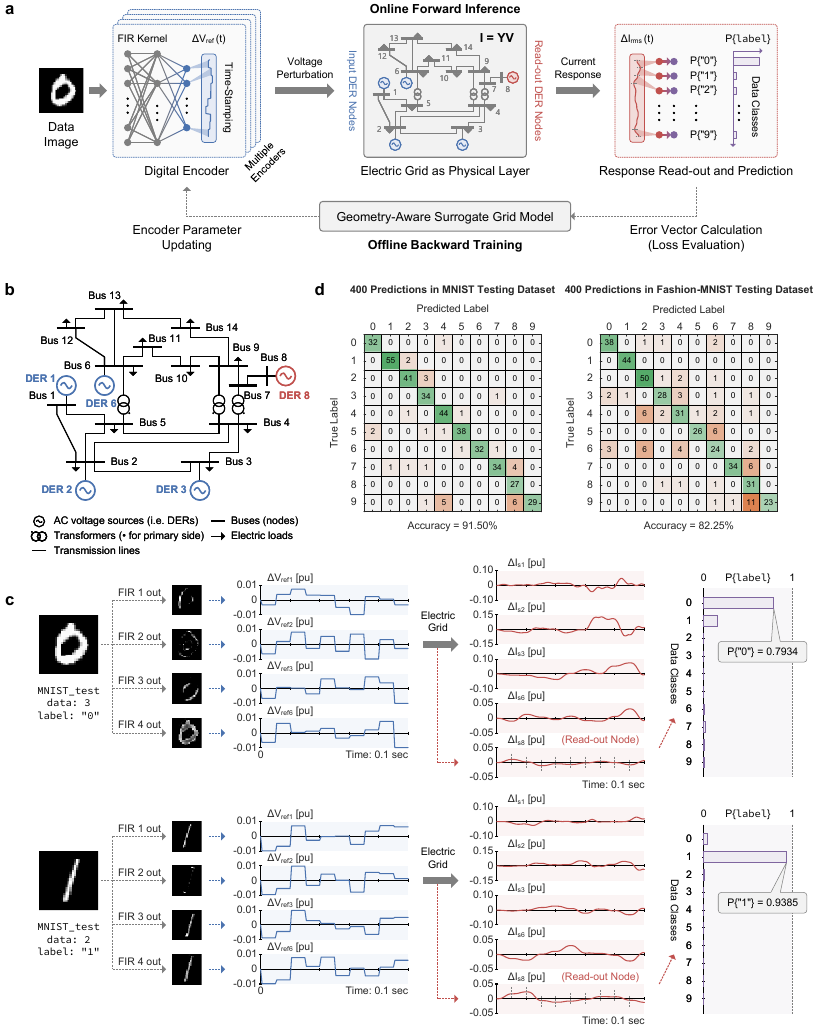}
\end{figure}
\clearpage
\refstepcounter{figure}\label{fig_1}
\noindent
\textbf{Figure \thefigure:} {\textbf{Electric power grids as physical computing substrates.} (a) Hybrid physical-digital computing architecture in which input data are encoded as voltage perturbations at converter-interfaced nodes, transformed by the physical substrate (the grid), and decoded from current responses, (b) IEEE 14-bus benchmark system used as the physical computing layer, showing the controllable input nodes and read-out node, (c) Representative voltage perturbations and resulting current responses for encoded inputs propagating through the physical network, (d) Classification performance on MNIST and Fashion-MNIST obtained from the physical computing pipeline. The results are obtained from the first 400 samples in each dataset, the encoders for MNIST and Fashion-MNIST datasets are trained for 2 epochs and 5 epochs, respectively, and the grid dynamics are simulated with Plexim PLECS \cite{plecs_manual}.} \\[30pt]

The IEEE 14-bus benchmark system \cite{IEEE14BusData}, shown in Fig.~\ref{fig_1}b, comprises 14 buses, 17 transmission lines, and 3 power transformers, with converter-interfaced DERs at nodes 1, 2, 3, 6, and 8. Four of these nodes provide controllable inputs, while node 8 is used as the principal read-out location. The input and read-out nodes are selected according to their electrical coupling through the network, providing heterogeneous sensitivities between input perturbations and the measured current responses.

Fig.~\ref{fig_1}c illustrates the encoding and the physical transformation for two representative test images from the MNIST dataset — the digit "0" and the digit "1". Each image is processed through four parallel FIR encoder channels, producing four distinct voltage perturbation waveforms $\mathrm{\Delta V_{ref1}(t)}$ through $\mathrm{\Delta V_{ref6}(t)}$ that are simultaneously applied to the input DERs 1, 2, 3 and 6. The waveforms differ systematically between the two digits, reflecting the different \textit{features} extracted by the FIR kernels from each image class. The physical grid responds to these voltage perturbations with current-response \textit{signatures} $\mathrm{\Delta I_{s1}(t)}$ through $\mathrm{\Delta I_{s8}(t)}$ at the monitored buses. For example, the digit "0" produces a current response that the decoder assigns probability $\mathrm{P}$\{"0"\} = 0.7934, while the digit "1" produces a qualitatively different response pattern assigned $\mathrm{P}$\{"1"\} = 0.9385. This distinguishability emerges from the topology-dependent coupling between input and reading-out buses, which responds differently to the \textit{features} of each image class as encoded in the voltage perturbation waveforms.

The classification performance of the complete system is summarized in the confusion matrices of Fig.~\ref{fig_1}d, evaluated on 400 test images drawn from each dataset through PLECS time-domain simulations, where the pipeline achieves 91.50\% accuracy on MNIST across all ten digit classes and 82.25\% on Fashion-MNIST. For MNIST dataset, the strongest classification performance occurs on digits "\texttt{1}" and "\texttt{8}", while the largest overlap occurs between visually similar digit classes such as "\texttt{3}" and "\texttt{5}" — pattens consistent with the spatial similarity of the digits under the FIR encoding. Fashion-MNIST exhibits the expected increase in inter-class confusion arising from the greater visual similarity between clothing categories. Importantly, both results are obtained with the same physical layer and encoder architecture, differing only in the FIR kernel weights optimized for each dataset.

Training of the encoder is performed offline using a geometry-aware surrogate model of the IEEE 14-bus system, which replicates the input-output behavior of the physical layer in software. The surrogate model is constructed from the analytically derived signal transformation of the IEEE 14-bus system and serves as the differentiable forward model where encoder parameters are updated via standard backpropagation. This surrogate-based training approach introduces a simulation-to-physical gap: it achieves 95.25\% accuracy on MNIST under ideal conditions, establishing an upper bound on the performance achievable through the admittance matrix transformation, while a 3.75\% difference is also observed compared with 91.5\% for PLECS time-domain simulation of the network. The origin of this difference is examined below through controlled perturbations of the physical operating conditions.

\subsection*{Network Topology Provides a Task-Independent Computational Prior}

The structure of the physical operator can be examined through the Jacobian matrix 
\begin{equation}
    \mathbf{J}
    =
    \frac{\partial \mathbf{I}_\mathrm{rms}}{\partial \mathbf{V}_\mathrm{ref}}
\end{equation}
which characterizes the sensitivity of the measured current responses to voltage perturbations applied at the controllable DER nodes. The Jacobian mapped onto the IEEE 14-bus topology is shown in Fig.~\ref{fig_2}a, mapped against the full topology of the IEEE 14-bus system across all 17 transmission lines and 3 power transformers. Its entries exhibit a sparse, structured pattern that follows the electrical connectivity of the network: Input nodes connected to the read-out through short, low-impedance paths exhibit larger sensitivities, whereas electrically distant or weakly coupled paths contribute smaller responses. Physically, KCL and Ohm's law restrict the admittance matrix to transformations consistent with network topology and current conservation, concentrating computational capacity along electrically meaningful coupling paths before classification training takes place.

This topology-dependent structure provides an inductive bias: relationships between inputs and outputs that would otherwise have to be learned from classification data are partially specified by the network itself. To quantify the computational consequence of this constraint, we construct a controlled comparison in which the FIR encoder, voltage-mapping stages, and output calibration are kept identical, while only the core computational operator is changed (Fig.~\ref{fig_2}c-e). In the physical system, the core operator is represented by a 933-parameter surrogate model of the IEEE 14-bus admittance transformation, fitted once to reproduce the grid input-output behavior and then frozen throughout classification training. In the digital baselines, the same operator is replaced by fully connected layers whose parameters are optimized directly for classification.

In Fig.~\ref{fig_2}c, the grid-based physical operator achieves a mean per-class MNIST accuracy of 0.915. For comparison, a single layer mapping 400 inputs to 10 outputs achieves 0.800 mean accuracy, while  two-layer fully connected networks with hidden widths of 20, 30, and 40 achieve 0.875, 0.902, and 0.970, respectively. The 30-hidden-unit digital transformation provides the closest accuracy match to the physical system and requires 12,340 task-trained parameters: 12,030 parameters in the 400-to-30 layer and 310 in the 30-to-10 layer. The physical operator is represented by 933 parameters, corresponding to a 13.2-fold reduction in the number of parameters used to represent the core computational transformation (\textit{\textbf{Remark:} This comparison should be interpreted as an operator-representation comparison rather than as a direct measurement of hardware parameter count or energy consumption}). The 933 surrogate parameters were not optimized for MNIST classification; instead, they describe the fixed physical admittance operator representing the physical substrate (the electric power grid). The comparison indicates the amount of task-trained parameterization required by the particular fully connected replacements evaluated here to reach similar classification performance. It does not establish a universal parameter advantage of electrical-network operators, nor does it measure end-to-end hardware complexity, energy use or inference latency, but validates the computational structure identified as dominated by the topology constraints collectively imposed by the conservation laws.

\clearpage
\begin{figure}[p]
    \centering
    \includegraphics[width=\linewidth]{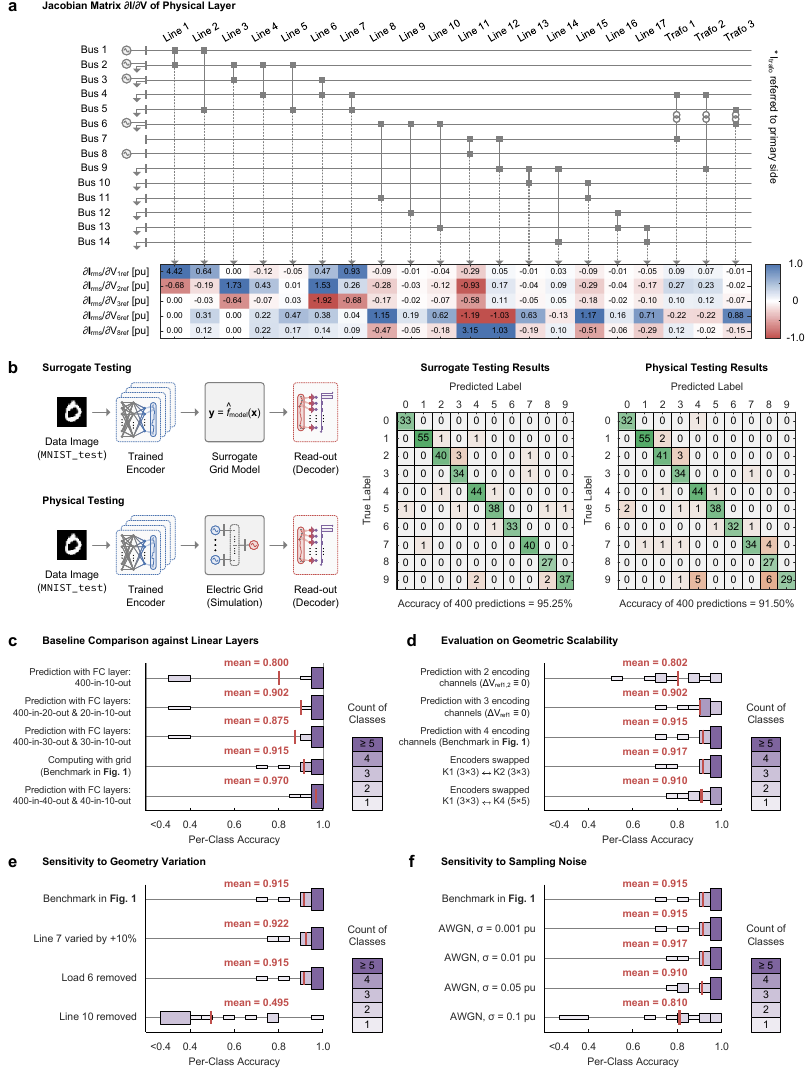}
\end{figure}
\clearpage
\refstepcounter{figure}\label{fig_2}
\noindent
\textbf{Figure \thefigure:} {\textbf{Topology-dependent physical operator and sensitivity analyses.} \textbf{a}, Jacobian matrix of the physical operator that maps the input voltage perturbations ($\Delta \mathrm{V}_\mathrm{ref}$) to line and transformer RMS current variations ($\Delta \mathrm{I}_\mathrm{rms} ^\mathrm{(line/trafo)}$). \textbf{b}, Evaluation tracks used in this study: a differentiable surrogate model supports encoder training and surrogate-model evaluation, whereas physical system replica realized via, e.g., PLECS time-domain simulations evaluates the trained encoder against the network model. \textbf{c}, Baseline comparison of the fixed network operator against task-trained fully-connected (FC) linear transformations under matched encoder, voltage-mapping and output-calibration settings, trained by 2 epochs. \textbf{d}, Sensitivity to the number of accessible encoding channels and to selected encoder-kernel assignments. \textbf{e}, Sensitivity to selected network changes. \textbf{f}, Sensitivity to sampling noise, where the read-out is performed subject to additive white Gaussian noise (AWGN). In \textbf{c}-\textbf{f}, per-class accuracy are evaluated under the \textit{physical testing} illustrated in \textbf{b} using the trained encoder and PLECS time-domain simulations with the first 400 samples in the MNIST testing dataset.} \\[30pt]

In Fig.~\ref{fig_2}d, we further examine whether the computational behavior depends strongly on the particular encoder configuration. Reducing the number of active encoding channels from four to three ($\mathrm{\Delta V_{ref1} \equiv 0}$) decreases mean accuracy marginally from 0.915 to 0.910, whereas reducing to two channels ($\mathrm{\Delta V_{ref1,2} \equiv 0}$) decreases it to 0.802. Swapping encoder kernels between channels — substituting the vertical-edge kernel of channel 1 with the cross-detector kernel of channel 2, or with the Laplacian kernel of channel 4 — produces accuracy of 0.917 and 0.910 respectively, close to the four-channel benchmark. The physical operator therefore supports a range of encoder configurations, while its performance degrades progressively as the dimensionality of the encoded input is reduced.

\subsection*{Physical Superposition Enables Multiplexed Inference}

The linearity of the physical transformation also permits multiple encoded inputs to share the same computational operator. For independent voltage perturbations,
\begin{equation}
    \mathbf{Y}(\Delta \mathbf{V}_1+\Delta \mathbf{V}_2)
    =
    \mathbf{Y}\Delta \mathbf{V}_1 + \mathbf{Y}\Delta \mathbf{V}_2,
\end{equation}
so their current contributions remain additive at the read-out. We test in the IEEE 14-bus surrogate model whether one fixed physical operator can be shared by multiple independently encoded inference streams without duplicating the modeled network transformation. The multiplexing results reported below are surrogate-model evaluations; they do not establish physical-hardware throughput or end-to-end energy savings.

\begin{figure}[ht]
    \centering
    \includegraphics[width=\linewidth]{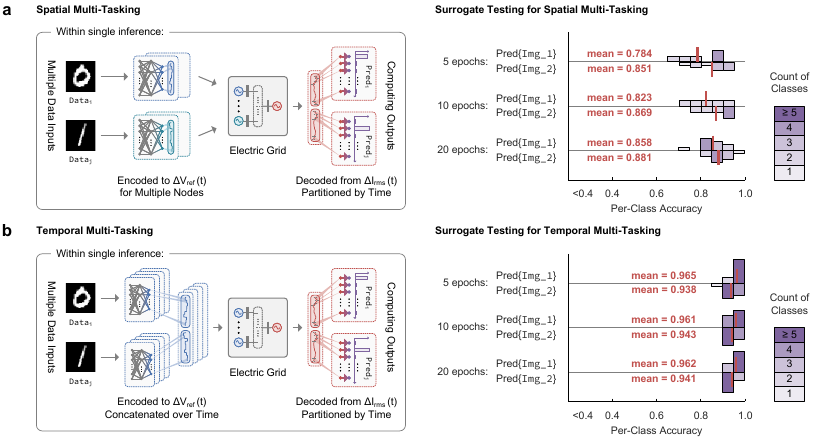}
    \caption{\textbf{Multiplexed inference enabled by physical superposition: spatial sharing and temporal reuse of a fixed electrical operator.} \textbf{a}, Spatial multi-tasking across input DER nodes and exemplary surrogate testing performances. \textbf{b}, Temporal multi-tasking over encoding time windows and exemplary surrogate testing performances. The distributions of per-class accuracy are inspected and obtained by shuffling all 10000 samples in the MNIST testing dataset with different random generator seeds for 2 data inputs.}
    \label{fig_3}
\end{figure}

We demonstrate two distinct multi-tasking configurations in Fig.~\ref{fig_3}. In spatial multi-tasking (see Fig.~\ref{fig_3}a), two images are encoded simultaneously at different subsets of the controllable DER nodes, and their responses are then decoded from temporally partitioned current responses at the read-out bus. In temporal multi-tasking (Fig.~\ref{fig_3}b), two images are encoded sequentially in time on the same input nodes, and the resulting current responses are separated into non-overlapping time windows for independent decoding. Both configurations reuse the same physical infrastructure (electrical network) and the same physical operator.

Spatial multi-tasking exhibits a clear topology-induced trade-off. In Fig.~\ref{fig_3}a, at 5 epochs, the two tasks achieve mean per-task accuracies of 0.784 and 0.851, increasing to 0.858 and 0.881 at 20 epochs. These values are below the single-task benchmark of 0.915 because spatial partitioning distributes the available input nodes between the tasks, giving each task access to only a subset of the columns of the physical operator and reduced effective dimensionality for classification. Moreover, the input nodes are not fully equivalent: their coupling strengths to the read-out depend on their location in the network. The observed asymmetry between the two tasks therefore reflects the topology-dependent allocation of physical coupling capacity.

Temporal multiplexing provides each successive task with the complete set of input nodes and hence access to the full physical operator. In Fig.~\ref{fig_3}b, the two tasks achieve mean accuracies of 0.965 and 0.938 after 5 epochs, and 0.962 and 0.941 after 20 epochs. Each task is encoded in a separate time window and decoded when the network returns to a point where the adjacent decoding windows are sufficiently separable. The higher accuracy relative to spatial multiplexing is consequently consistent with the complete reuse of the physical operator rather than with a reduction of its available spatial couplings. These tests demonstrate time-division reuse of a fixed physical computational operator alongside spatial sharing of the same infrastructure.

Spatial sharing and temporal reuse arise from different uses of the same fixed operator. Spatial sharing exploits simultaneous perturbations applied at distinct node groups and incurs a reduction in the coupling channels available to each stream. Temporal reuse assigns successive inputs to low-to-non-overlapping windows and gives each stream access to the complete operator. The present experiments demonstrate these two modes in a surrogate model: They do not establish zero crosstalk, intrinsic temporal memory, or a throughput advantage relative to independent digital execution.

\subsection*{Computational Utility Persists within a Finite Operating Envelope}

A computational operator embedded in an operating power network must retain its task-relevant transformation as the electrical parameters vary. We therefore perturb line impedances, loads, and measurement noise while keeping the classification pipeline unchanged. These experiments characterize the range over which the tested physical operator remains usable rather than establishing robustness over all possible grid operating conditions.

Fig.~\ref{fig_2}e shows the response to changes in network geometry. A 10\% increase in the impedance of line 7 changes the mean accuracy only slightly, from 0.915 to 0.922. Removing load 6 leaves the accuracy unchanged at 0.915, indicating that changes in the passive load at this location do not substantially alter the input-to-read-out coupling used by the classifier. In contrast, removing line 10 reduces the mean accuracy to 0.495, demonstrating that disruption of a critical electrical path can strongly alter the physical computational operator. The different responses to line and load perturbations are therefore consistent with the topology dependence identified from the Jacobian.

We then evaluate additive white Gaussian noise on the sampled current responses in Fig.~\ref{fig_2}f. Across $\sigma=0.001$--$0.05$~pu, the mean accuracy remains within approximately one percentage point of the benchmark, varying from 0.917 to 0.910. At $\sigma=0.1$~pu, accuracy decreases to 0.810. The noise level around $\sigma=0.05$~pu produces a degradation comparable to the 3.75\% difference between the ideal surrogate and the time-domain physical simulation. In the PLECS simulation settings, effective perturbations arise from, e.g., numerical integration tolerances and measurement discretization. The comparable degradation at $\sigma\approx$ 0.05~pu indicates that read-out and simulation-related perturbations of this magnitude could account for a substantial part of the surrogate–simulation difference, although the present analysis does not uniquely identify individual contributions.

Together, these experiments define a finite computational operating envelope: moderate parameters and measurement perturbations preserve the task-relevant transformation in the tested system, whereas topological changes that remove important coupling paths can substantially degrade performance.

\subsection*{Computational Utility Depends on Input Representation}

\begin{figure}[ht]
    \centering
    \includegraphics[width=0.85\linewidth]{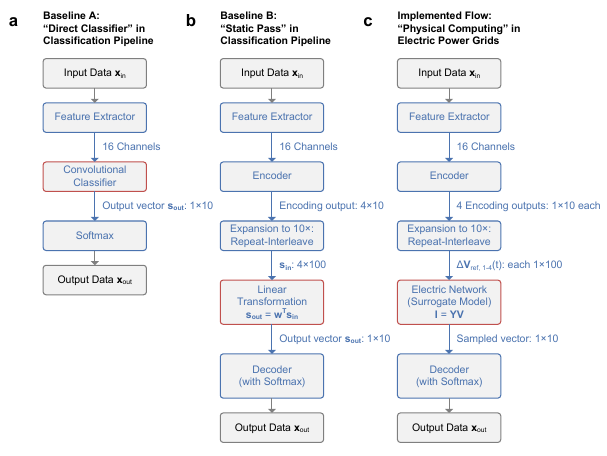}
    \caption{\textbf{Classification pipelines that are compared for evaluating the capability of generalizing the physical computing infrastructure for more complex datasets.}}
    \label{fig_5_new}
\end{figure}

The contribution of the physical operator should depend on the structure of the representation presented to it. To isolate this contribution, we compare three otherwise related classification pipelines (Fig.~\ref{fig_5_new}): a digital \textit{direct classifier} operating on the extracted features, a \textit{static pass} that receives the same encoded signals but bypasses the physical network, and the proposed \textit{physical computing} pathway. The \textit{direct classifier} tests whether the extracted representation contains sufficient information for classification by itself, whereas the \textit{static pass} provides the critical control for determining whether the specific physical transformation contributes beyond a learned linear combination of the same encoded features.

\begin{figure}[ht]
    \centering
    \includegraphics[width=\linewidth]{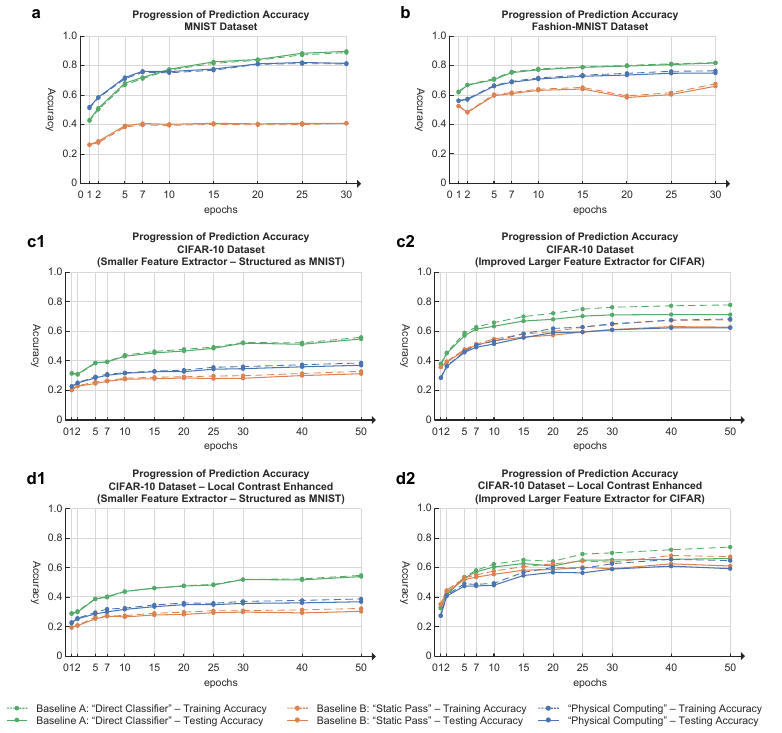}
    \caption{\textbf{Dependence of physical computational utility on input representation.} \textbf{a}, Evaluation on MNIST dataset. \textbf{b}, Evaluation on Fashion-MNIST dataset. \textbf{c} and \textbf{d}, Evaluation on CIFAR-10 dataset (\textbf{c}, without pre-processing, and \textbf{d}, with local contrast enhanced).}
    \label{fig_6_new}
\end{figure}

For the MNIST and Fashion-MNIST datasets, the physical pathway consistently outperforms the static pass during training and testing while approaching the performance of the direct classifier in Fig.~\ref{fig_6_new}a,b. Hence, the current responses generated by the grid provide a task-relevant transformation beyond a direct linear recombination of the encoded signals. This result complements the controlled parameter comparison above: the physical operator is useful since its fixed topology-dependent transformation changes the representation presented to the decoder.

We then increase the input complexity using CIFAR-10 dataset \cite{cifar10}. The same compact feature-extraction architecture used for the grayscale datasets is first adapted from one to three input channels to process the RGB images in Fig.~\ref{fig_6_new}c. The resulting representation is more difficult to classify, and all three pipelines exhibit lower accuracy than on MNIST/Fashion-MNIST. We then evaluate an increased-capacity feature extractor together with local contrast enhancement using Contrast Limited Adaptive Histogram Equalization (CLAHE) in Fig.~\ref{fig_6_new}d. The stronger upstream feature extraction improves the separability of the encoded representation and reduces the difference between the physical and static pathways.

This dependence on upstream representation is important for interpreting the role of the physical substrate. Since the grid transformation itself is linear, additional nonlinear processing before the grid can make the features increasingly linearly separable and consequently reduce the incremental benefit of the physical operator. Conversely, when the encoded representation retains task-relevant structure that is not readily separable by a direct linear mapping, the topology-dependent grid transformation provides a larger contribution.

To extend this analysis beyond the ten-class image datasets, we further evaluate the physical pathway on the Butterflies-and-Moths dataset (see \textit{Supplementary Note S1}), which contains 13,094 training images and 500 test images distributed over 100 species. Following the supplementary evaluation protocol, 25 independent experiments are performed, with 10 species selected at random for each experiment. Both the compact feature extractor used for MNIST/Fashion-MNIST and the deeper feature extractor used for CIFAR-10 are evaluated, with additional tests using local-contrast enhancement and background removal. These experiments extend the representation-dependence analysis to natural-image data drawn from a larger 100-species collection while retaining a ten-class classification problem in each trial. They do not constitute a single 100-class classification evaluation; rather, performance is averaged over 25 independently sampled ten-species subsets using the same underlying surrogate operator.

Overall, the complex-dataset experiments show that the physical substrate does not replace digital feature extraction. Instead, it provides a fixed, task-independent transformation whose usefulness depends on the structure of the representation presented to it. The static-pass comparison therefore provides a direct measure of when the physical network adds task-relevant
computation beyond the encoded representation alone.

\section*{Discussions}

The 13.2-fold difference in represented parameter count between the fitted surrogate operator and the accuracy-matched fully connected core is consistent with a broader principle in physics-informed computation: physical laws can encode prior information about input-output relationships that would otherwise need to be learned from data \cite{karniadakis2021physics}. In constrained physical systems, this information can reduce the dimensionality of the task-relevant hypothesis space and thereby reduce the number of free parameters required by a corresponding unconstrained digital representation \cite{raissi2019physics}. KCL and Ohm's law impose particularly strong constraints on the electric-grid operator. They determine the admittance relationships from network topology and line impedances rather than allowing the entries of the computational operator to be chosen independently. For the IEEE 14-bus system, the accessible physical couplings are therefore determined by a sparse network containing 17 transmission lines and 3 power transformers, whereas the fully connected digital replacement is free to learn its coupling structure directly from classification data. The resulting comparison provides a quantitative example of how a physical law can act as a task-independent computational prior: the fixed grid operator reaches a mean accuracy comparable to the 12,340-parameter digital replacement while its fitted surrogate representation contains 933 parameters that identify the simulated electrical response rather than being trained on the classification task.

This interpretation is consistent with recent physics-informed graph neural networks in which conservation laws are embedded as architectural constraints to reduce the need to learn physical interaction structure from data \cite{sharma2026physics}. The analogy is useful, but the present system differs in an important respect. In physics-informed digital networks, the physical constraints are deliberately incorporated into a computational architecture; here, the constraints arise from an infrastructure whose primary purpose is energy delivery. The computational operator is therefore not designed from the task backward. Rather, it is inherited from the network topology and governing equations. This distinction is particularly relevant to unconventional computing because it shifts the design principle from engineering physical hardware to implement a desired operator toward identifying useful operators that are already present in electric power grids infrastructure.

This work is distinct from conventional neuromorphic computing in both principle and implementation \cite{yan2024emerging, kudithipudi2025neuromorphic, schuman2022neuromorphic}. Resistive crossbar architectures in neuromorphic computing, for example, exploit Ohm's law and Kirchhoff's current law to implement matrix-vector multiplication, but their conductance elements are deliberately programmed to represent task-specific weights \cite{9141337,deng2026memristive}. The physical network is therefore configured around the computational objective, and the conductance matrix can in principle be optimized to approximate a broad class of desired weight matrices \cite{chen2020wafer, dalgaty2024mosaic, ambrogio2018equivalent}. In the electric grid considered here, the corresponding admittance matrix is fixed by infrastructure built for power delivery: no element of $\mathbf{Y}$ is assigned a classification weight, and the network does not provide arbitrary element-wise weight programming. The same physical constraints that limit the expressivity of the grid therefore provide its computational prior. Rather than competing with programmable neuromorphic hardware on generality, infrastructure-native computing exploits constrained physical operators for tasks for which their structure is useful.

Its distinction extends to other physical neural computing substrates. Wave-based systems based on acoustic cavities, microwave structures, and optical propagation can exploit nonlinear mixing, interference, scattering, or complex dynamical responses as computational mechanisms \cite{stern2023learning, iten2020discovering}. The grid considered here instead provides a structured and interpretable transformation. Its steady-state operator is analytically related to network topology and line impedances \cite{song2025unlocking, sahoo2025spike} and can be characterized experimentally through the Jacobian $\partial \mathbf{I}_\mathrm{rms}/\partial \mathbf{V}_\mathrm{ref}$. This provides a direct link between physical connectivity and computational coupling: the operator can be interpreted from the network geometry rather than treated solely as a black-box physical response. This interpretability also suggests a different route to physical-computing design, in which computational properties can potentially be influenced by modifying network topology or operating conditions rather than by independently programming every computational weight.

A further distinction is that the proposed computational substrate is embedded in infrastructure that already contains power electronic interfaces through which controlled voltage perturbations can be applied. The $\pm0.01$~pu perturbations used here lie within the voltage-flexibility range discussed for converter-interfaced grid resources \cite{entsoe2024report, entsoe2025report}. This indicates that the physical signals required by the present computational demonstration are compatible, in magnitude, with control actions already considered in grid operation. It does not by itself establish that computation can be deployed as an ancillary service or that no additional sensing, protection, communication, or control infrastructure would be required. Rather, it identifies a plausible deployment pathway in which computation is superimposed on existing converter controllability without changing the underlying network topology.

The experiments with increasingly complex datasets further clarify when the physical operator is computationally useful. The comparison between the physical pathway and the static-pass baseline in Fig.~\ref{fig_6_new} shows that the grid contributes a task-relevant transformation beyond a learned linear recombination of the same encoded features. However, this contribution is not independent of the input representation. For MNIST and Fashion-MNIST, the physical pathway substantially outperforms the static pass, whereas for CIFAR-10 the difference becomes smaller as the upstream feature extractor is made more expressive and local contrast enhancement improves feature separability. The Butterflies-and-Moths experiments extend this observation to a 100-species classification setting. These results suggest that the value of an infrastructure-native operator is determined jointly by the structure of the physical coupling and by the representation presented to it: the physical substrate is most useful when task-relevant structure remains that cannot be captured adequately by a direct linear mapping. Conversely, as upstream learning makes the representation increasingly linearly separable, the incremental contribution of the fixed physical operator diminishes. This representation dependence is important because it defines a more realistic role for infrastructure-native computing: not as a replacement for learned digital feature extraction, but as a physical transformation that can complement it.

The superposition experiments in Fig.~\ref{fig_3} provide a second computational property that follows directly from the linearity of the physical operator. Independent inputs can be combined through the same electrical network and subsequently separated spatially or temporally through their encoded channels. Spatial multiplexing demonstrates the constraint imposed by finite controllable nodes, with mean per-task accuracies of 0.858 and 0.881 after 20 epochs, whereas temporal multiplexing allows each task to reuse the complete operator and reaches 0.962 and 0.941. The contrast between these two modes shows that the computational capacity of infrastructure-native systems is not determined solely by the size of the physical network, but also by how controllable degrees of freedom and temporal bandwidth are allocated among tasks. This provides a potential route toward sharing a fixed physical operator across multiple inference streams without replicating its physical weights.

Several limitations define the next stage of the work. First, training is performed offline through a geometry-aware surrogate model of the grid, introducing a simulation-to-reality modeling gap that is characterized but not eliminated in the present study. \textit{In-situ} training directly through physical measurements could remove the surrogate as an intermediary and allow the encoder to adapt to the actual physical input-output response, analogous to backpropagation-free training approaches developed for other physical neural substrates \cite{momeni2023}. Second, the steady-state physical operator considered here is linear. Extending the architecture to multiple cascaded physical layers with trainable nonlinear transformations would require nonlinear physical mechanisms, potentially including the switching and saturation characteristics of power-electronic converters. Such an extension is conceptually distinct from the present demonstration and remains to be established experimentally. Third, the computation delays are essentially attributed to the time window where the encoded perturbations take place. While longer windows in turn benefit the read-out resolution where overlapping in dynamics is mitigated, a quantification of such trade-off remains to be observed. Fourth, all results in this study are obtained with the IEEE 14-bus benchmark system and time-domain simulation. Physical validation on a laboratory-scale grid with programmable converters would therefore be an important next step for assessing measurement bandwidth, control latency, converter interactions, protection constraints, and hardware-specific variability \cite{xu2026physical}.

Finally, the scaling of computational capacity with network size remains an open empirical question. For sparse grid topologies in which the number of branches scales approximately with the number of buses, the number of independent physical parameters determining the network operator can scale approximately as $\mathrm{O}(\mathrm{N})$, whereas a fully connected digital representation of comparable dimensionality contains $\mathrm{O}(\mathrm{N}^2)$ free parameters \cite{dorfler2012kron}. The present 14-bus comparison is consistent with this structural difference, but it does not establish that the measured 13.2-fold advantage will grow monotonically with system size. Larger benchmark networks, such as the IEEE 33-bus \cite{baran2002network} and IEEE 118-bus systems \cite{5491276}, provide tractable test cases for determining whether operator representation, controllability, multiplexing capacity, and task performance follow predictable scaling relationships. Establishing such scaling laws would determine whether infrastructure-native computation can progress from a single physical operator to distributed computation across larger electric networks.

\clearpage

\printbibliography

\clearpage
\setcounter{page}{1}
\renewcommand{\thefigure}{S\arabic{figure}}
\renewcommand{\thetable}{S\arabic{table}}
\renewcommand{\theequation}{S\arabic{equation}}
\renewcommand{\thepage}{S\arabic{page}}

\setcounter{figure}{0}
\setcounter{table}{0}
\setcounter{equation}{0}

\section*{Materials and Methods}

\subsection*{Specifications of the IEEE Benchmark Grid}

The study case is defined based on the IEEE 14-bus benchmark system (raw data from \cite{IEEE14BusData}, which is a three-phase AC microgrid as illustrated in Fig.~\ref{fig_ieee14bus}). The configurations of the study case are given as follows. Key parameter ratings include: nominal voltage $\mathrm{V_{base}} = 138\;\mathrm{kV}$, nominal power (three-phase) $\mathrm{S_{base}} = 100\;\mathrm{MVA}$, and fundamental frequency $\mathrm{f_{1}} = 50\;\mathrm{Hz}$. To simplify the analysis, other conditions that not specified in this section (sources/line power limits) are disregarded in the study case in this article.

\begin{itemize}
    \item Buses 1, 2, 3, 6, 8 are connected to generators/power sources, which are regarded as distributed energy resources (DERs) interfaced by power electronic converters. These buses are denoted as \textit{active buses} in contrast to the rest buses as \textit{passive buses} that are not connected to generators/power sources. It is assumed that the magnitude $\mathrm{V}$ (specified in root-mean-square (RMS) values) and phases $\mathrm{\theta}$ of the voltages of these buses are ideally regulated neglecting source dynamics, of which the default values are given in Table~\ref{tab_DER}. In this work, Buses 1, 2, 3, 6 are defined as input DER nodes, while Bus 8 is defined as output DER nodes.

    \item Buses 2, 3, 4, 5, 6, 9, 10, 11, 12, 13, 14 are connected to loads locally. For simplicity, the loads are assumed to be configured as three-phase \textit{RL} series (except for the load at Bus 4 configured as \textit{RC} series due to its negative reactive power rating). The power ratings under nominal conditions are given in Table~\ref{tab_load}.

    \item Transmission lines connecting the buses are also modeled as \textit{RL} series. The parameters are specified as equivalent serial resistances $\mathrm{R}$ and reactances $\mathrm{X}$ under nominal conditions in Table~\ref{tab_line}.

    \item The shunt capacitances of some transmission lines are also encompassed and specified as equivalent total susceptances $\mathrm{B}$ under nominal conditions in Table~\ref{tab_shunt}. In the case study in this article, the $\pi$ model is applied to these transmission lines with shunt capacitances, i.e., the susceptance of each line is modeled as two equivalent capacitances equally distributed to both terminals.

    \item Three transformers are considered in the system, which are installed between Buses 4 \& 7, 4 \& 9, and 5 \& 6, respectively. The turn ratios $\mathrm{\alpha}$ ($\mathrm{N_2/N_1}$, where $\mathrm{N_1}$ and $\mathrm{N_2}$ are the turns on primary and secondary sides, respectively) and equivalent reactances $\mathrm{X}$ (all referred to primary sides) are given in Table~\ref{tab_line}.
\end{itemize}

\subsection*{Exemplary Computation Task}

A typical image classification task is demonstrated on the benchmark grid showcasing the ability to compute, where the MNIST and Fashion-MNIST datasets are designated as the inputs. The MNIST dataset contains grayscale images of handwritten digits, and the Fashion-MNIST dataset contains grayscale images of Zalando's article images.

MNIST and Fashion-MNIST datasets both consist of 60,000 images in the training set and 10,000 in the testing set. In MNIST \cite{mnist} and Fashion-MNIST datasets \cite{fashion-mnist}, each image contains $28 \times 28 = 784$ pixels and is labeled with an integer ranging from 0 to 9 (label descriptions given in Table~\ref{tab_dataset}, corresponding to 10 classes of handwritten digits or fashion items, respectively). In the dataset files, the pixels are stored as 8-bit grayscale intensities (0–255), flattened into a one-dimensional array in row-major order from the top-left corner.

\subsection*{Specifications of the Encoder and Decoder}

The finite-impulse-response (FIR) encoder first extracts features from the input image $\mathbf{x}_{28\times28}$ using parallel convolutional layers with different kernel configurations followed by \texttt{ReLU} activation. The resulting feature maps are flattened and linearly projected into $10\times1$ vectors with normalization. Each vector element is then temporally expanded by repeating over 10 consecutive samples, corresponding to a sampling frequency of 1 kHz, thereby constructing the time-stamped signals $\mathrm{\Delta}\mathrm{V}_\mathrm{ref}$ as perturbations to bus voltage references that are passed to the electric grid.

\begin{figure}[ht]
    \centering
    \includegraphics[width=0.6\linewidth]{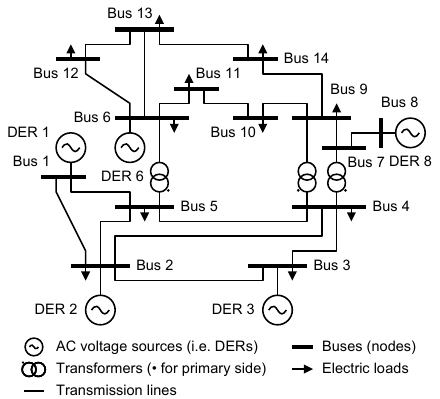}
    \caption{IEEE 14-bus benchmark system (an AC system).}
    \label{fig_ieee14bus}
\end{figure}

In this article, the voltage perturbations $\mathrm{\Delta}\mathrm{V}_\mathrm{ref}$ occur at DER 1, DER 2, DER 3 and DER 6 of Fig.~\ref{fig_ieee14bus} unless otherwise specified. 4 different FIR kernels are deployed and respectively initialized as follows:

\begin{equation}
\label{eq_ker_init}
\begin{aligned}
    \mathtt{Kernel\_1\:}_{3\times3} &:=
    \begin{bmatrix}
        -1 & 0 & 1 \\
        -2 & 0 & 2 \\
        -1 & 0 & 1
    \end{bmatrix}
    \qquad
    &
    \mathtt{Kernel\_2\:}_{3\times3} &:=
    \begin{bmatrix}
        0 &  1 & 0 \\
        1 & -4 & 1 \\
        0 &  1 & 0
    \end{bmatrix}
    \\[1em]
    \mathtt{Kernel\_3\:}_{3\times3} &:=
    \begin{bmatrix}
        2 &  1 &  0 \\
        1 &  0 & -1 \\
        0 & -1 & -2
    \end{bmatrix}
    \qquad
    &
    \mathtt{Kernel\_4\:}_{5\times5} &:=
    \begin{bmatrix}
         0 &  0 & -1 &  0 &  0 \\
         0 & -1 & -2 & -1 &  0 \\
        -1 & -2 & 16 & -2 & -1 \\
         0 & -1 & -2 & -1 &  0 \\
         0 &  0 & -1 &  0 &  0
    \end{bmatrix}
\end{aligned}
\end{equation}
which highlight the vertical edges, crosses, corners/endpoints and loops in the input images, respectively. Subsequently, the linear projection following the FIR convolution layers are configured with input/output dimensions of: 676/10 for the branches associated with Kernels 1-3, while 576/10 for the branches associated with Kernel 4.

The root-mean-square (RMS) current variation at DER 8 is sampled over a time period of 0.1 s with a sampling frequency of 1 kHz. The output sequence is defined as the time-domain integral of the RMS current variation samples.

The decoder first divides the output sequence into 10 time segments and samples the output sequence at every 10\textsuperscript{th} element. The increments between adjacent elements are then calculated to account for the accumulated current variation within individual segments. The resulting $10\times1$ vector is subsequently transformed using learnable scaling and offset parameters (1 gain and 1 bias assigned to each segment), thereby mapping the output into 10 classification bins. Finally, a \textit{softmax} operation is applied to obtain normalized class likelihoods.

The encoders/decoders are implemented and trained with PyTorch framework. Unless otherwise specified, the encoders/decoders are trained using Adam optimizer with following hyperparameters by default: learning rate 10\textsuperscript{-3}, batch size 16 and random seed 42.

\subsection*{Implementation of the Surrogate Grid Model}

To achieve high parameter efficiency, the forward pass of the surrogate grid model is implemented as a physics-informed dynamical mapping from voltage perturbations to the integral of current responses. The model emulates the spatial and temporal propagation and superposition behaviors of the electric network through coupled linear projections and causal convolutions.

Let the input sequence be denoted as $\mathbf{x}\mathrm{(t)} = [\mathrm{x_1(t)}, \mathrm{x_2(t)}, ..., \mathrm{x_n(t)}]^\mathrm{T}$, where $\mathrm{n}$ is the number of input DER nodes, then for each $\mathrm{i \in \{1, 2, ..., n\}}$, the inputs $\mathrm{x_i(t)}$ are first projected onto two components with learnable coefficients $\mathrm{u_i}$ mimicking the rotating reference frame (\textit{dq} frame, where \textit{d} stands for \textit{direct} component in direct alignment with the \textit{abc} reference vector and \textit{q} stands for \textit{quadrature} component at a quadrature angle to the direct component) decomposition in three-phase electric circuits:
\begin{equation}
\label{eq_surr_1}
\begin{aligned}
    \mathrm{x_i^\mathrm{d}(t)} &= \mathrm{u_i^\mathrm{d}x_i^\mathrm{d}(t)} \\
    \mathrm{x_i^\mathrm{q}(t)} &= \mathrm{u_i^\mathrm{q}x_i^\mathrm{q}(t)}
\end{aligned}
\end{equation}

The instantaneous feedforward responses $\mathbf{x}_\mathrm{ff}$ are then obtained through weighted superposition with learnable coefficients $\mathbf{W}$ and $\mathrm{b_{ff}}$:
\begin{equation}
\label{eq_surr_2}
\begin{aligned}
    \mathbf{x}_\mathrm{ff}^\mathrm{d}\mathrm{(t)} &= \mathbf{W}^\mathrm{dd}\mathbf{x}^\mathrm{d}\mathrm{(t)} + \mathbf{W}^\mathrm{qd}\mathbf{x}^\mathrm{q}\mathrm{(t)} + \mathrm{b_{ff}^d} \\
    \mathbf{x}_\mathrm{ff}^\mathrm{q}\mathrm{(t)} &= \mathbf{W}^\mathrm{dq}\mathbf{x}^\mathrm{d}\mathrm{(t)} + \mathbf{W}^\mathrm{qq}\mathbf{x}^\mathrm{q}\mathrm{(t)} + \mathrm{b_{ff}^q}
\end{aligned}
\end{equation}
followed by causal dynamics:
\begin{equation}
\label{eq_surr_3}
\begin{aligned}
    \mathbf{x}_\mathrm{dyn}^\mathrm{d}\mathrm{(t)} &= \mathbf{x}_\mathrm{ff}^\mathrm{d}\mathrm{(t)} + (\mathbf{G}^\mathrm{dd}*\mathbf{x}_\mathrm{ff}^\mathrm{d})\mathrm{(t)} + (\mathbf{G}^\mathrm{dq}*\mathbf{x}_\mathrm{ff}^\mathrm{q})\mathrm{(t)} + \mathrm{b_{dyn}^d} \\
    \mathbf{x}_\mathrm{dyn}^\mathrm{q}\mathrm{(t)} &= \mathbf{x}_\mathrm{ff}^\mathrm{q}\mathrm{(t)} + (\mathbf{G}^\mathrm{qd}*\mathbf{x}_\mathrm{ff}^\mathrm{d})\mathrm{(t)} + (\mathbf{G}^\mathrm{qq}*\mathbf{x}_\mathrm{ff}^\mathrm{q})\mathrm{(t)} + \mathrm{b_{dyn}^q}
\end{aligned}
\end{equation}
where "$*$" denotes casual convolution, and $\mathbf{G}$ and $\mathrm{b_{dyn}}$ are learnable kernels and bias, respectively. These convolutions represent the forward dynamics and coupling induced by the electric network.

The resulting states representing current responses are then calculated through recursive dynamics and coupling as another convolutional operation with learnable kernels $\mathbf{H}$ and bias $\mathrm{b_{y}}$, featuring the temporal memory effects of the electric network:
\begin{equation}
\label{eq_surr_4}
\begin{aligned}
    \mathbf{y}^\mathrm{d}\mathrm{(t)} &= \mathbf{x}_\mathrm{dyn}^\mathrm{d}\mathrm{(t)} + (\mathbf{H}^\mathrm{dd}*\mathbf{x}_\mathrm{dyn}^\mathrm{d})\mathrm{(t)} + (\mathbf{H}^\mathrm{dq}*\mathbf{x}_\mathrm{dyn}^\mathrm{q})\mathrm{(t)} + \mathrm{b_{y}^d} \\
    \mathbf{y}^\mathrm{q}\mathrm{(t)} &= \mathbf{x}_\mathrm{dyn}^\mathrm{q}\mathrm{(t)} + (\mathbf{H}^\mathrm{qd}*\mathbf{x}_\mathrm{dyn}^\mathrm{d})\mathrm{(t)} + (\mathbf{H}^\mathrm{qq}*\mathbf{x}_\mathrm{dyn}^\mathrm{q})\mathrm{(t)} + \mathrm{b_{y}^q}
\end{aligned}
\end{equation}

The combined RMS-like response $\mathbf{y}\mathrm{(t)}$ is then computed as:
\begin{equation}
\label{eq_surr_5}
    \mathbf{y}\mathrm{(t)} = \mathrm{\nu\sqrt{(\mathbf{y}^\mathrm{d})^2\mathrm{(t)}+(\mathbf{y}^\mathrm{q})^2\mathrm{(t)}+\epsilon}+c}
\end{equation}
where $\mathrm{\nu}$ and $\mathrm{c}$ are learnable coefficients, and $\mathrm{\epsilon}$ is a minimal constant for numerical stabilization.

Finally, the output $\mathbf{z}\mathrm{(t)}$ corresponding to accumulated current variation is obtained through an additional causal aggregation process with learnable kernels $\mathbf{p}$ and bias $\mathrm{b_{z}}$:
\begin{equation}
\label{eq_surr_6}
    \mathbf{z}\mathrm{(t)} = (\mathbf{p}*\mathbf{y})\mathrm{(t)} + \mathrm{b_{z}} = \mathrm{\sum_{\tau \leq t}{p(\tau)y(t-\tau)} + b_{z}}
\end{equation}

To better represent the dynamics of the target electric network, the surrogate grid model is trained using a hybrid loss function that jointly constrains signal fidelity, temporal dynamics and energy consistency. The overall loss function is defined as:
\begin{equation}
\label{eq_surr_l1}
    \mathcal{L} = \mathcal{L}_\mathrm{mse} + \mathrm{\lambda_{sampl}}\mathcal{L}_\mathrm{sampl} + \mathrm{\lambda_{deriv}}\mathcal{L}_\mathrm{deriv} + \mathrm{\lambda_{energ}}\mathcal{L}_\mathrm{energ}
\end{equation}
where, $\mathrm{\lambda_{sampl}}$, $\mathrm{\lambda_{deriv}}$ and $\mathrm{\lambda_{energ}}$ are the weighting coefficients.

Let $\mathbf{z}_\mathrm{pred}$ and $\mathbf{z}_\mathrm{true}$ denote the predicted and target data sequences, respectively, the The primary objective is the mean squared error between $\mathbf{z}_\mathrm{pred}$ and $\mathbf{z}_\mathrm{true}$:
\begin{equation}
\label{eq_surr_l2}
    \mathcal{L}_\mathrm{mse} = \mathrm{MSE(\mathbf{z}_\mathrm{pred}, \mathbf{z}_\mathrm{true})}
\end{equation}

$\mathcal{L}_\mathrm{sampl}$ is designated to preserve the consistency in representative samples (every 10 samples) and the steady states:
\begin{equation}
\label{eq_surr_l3}
\begin{aligned}
    \mathcal{L}_\mathrm{sampl} = & \mathrm{MSE(\mathbf{z}_\mathrm{pred}(:, 0:10:N, :), \mathbf{z}_\mathrm{true}(:, 0:10:N, :))} + \\
    & \mathrm{MSE(\mathbf{z}_\mathrm{pred}(:, N, :), \mathbf{z}_\mathrm{true}(:, N, :))}
\end{aligned}
\end{equation}
where $\mathrm{N}$ represents the length (number of data samples) of each data sequence.

$\mathcal{L}_\mathrm{deriv}$ aims to preserve the temporal dynamics:
\begin{equation}
\label{eq_surr_l4}
    \mathcal{L}_\mathrm{deriv} = \mathrm{MSE(\Delta \mathbf{z}_\mathrm{pred}, \Delta \mathbf{z}_\mathrm{true})} + \mathrm{\lambda_{deriv}}\mathrm{MSE(\Delta^2 \mathbf{z}_\mathrm{pred}, \Delta^2 \mathbf{z}_\mathrm{true})}
\end{equation}
where, $\Delta \mathbf{z}$ and $\Delta^2 \mathbf{z}$ represent the first-order and second-order temporal increments of the data sequences, respectively ($\mathrm{T_s}$ being the sampling period of the sequences):
\begin{equation}
\label{eq_surr_l5}
\begin{aligned}
    \Delta \mathbf{z}\mathrm{(t)} &= \mathbf{z}\mathrm{(t)}-\mathbf{z}\mathrm{(t-T_s)} \\
    \Delta^2 \mathbf{z}\mathrm{(t)} &= \mathbf{z}\mathrm{(t)}-2\mathbf{z}\mathrm{(t-T_s)}+\mathbf{z}\mathrm{(t-2T_s)}
\end{aligned}
\end{equation}

$\mathcal{L}_\mathrm{energ}$ serves for energy consistency and penalizes discrepancies in signal energy:
\begin{equation}
\label{eq_surr_l6}
    \mathcal{L}_\mathrm{energ} = \mathrm{MSE(\mathbf{z}_\mathrm{pred}^2, \mathbf{z}_\mathrm{true}^2)}
\end{equation}

In this work, the weighting coefficients are set as:
\begin{equation}
\label{eq_surr_l7}
    \mathrm{\lambda_{sampl}} = 0.5,\quad \mathrm{\lambda_{deriv}} = 0.1,\quad \mathrm{\lambda_{energ}} = 0.5
\end{equation}

The surrogate grid model is implemented and trained with PyTorch framework. Unless otherwise specified, the model is trained using Adam optimizer with following hyperparameters by default: 100 epochs, learning rate 2\texttimes10\textsuperscript{-3}, batch size 1 and random seed 42 to ensure the reproducibility of results.

\subsection*{Data Acquisition}

The cases are performed with procedures generally consisting of: (1) Data acquisition through time-domain simulations for training the surrogate grid model; (2) Training of the encoder and decoder while keeping surrogate grid model fixed; (3) \textit{Surrogate testing} in Fig.~\ref{fig_2} using the trained surrogate grid model; (4) Evaluations through repeated PLECS time-domain simulations representing \textit{physical testing} in Fig.~\ref{fig_2}, where the voltage perturbations $\mathrm{\Delta {V}_{ref}}$ are generated from the trained encoders and injected into the physical grid simulations; (5) The resulting system responses $\mathrm{\Delta {I}_{rms}}$ in (3) or (4) are subsequently processed by the decoder, followed by a \textit{softmax} operation to obtain the normalized class likelihoods; (6) The class with the highest likelihood is identified as the classification result, and the resulting classification statistics are used to evaluate the computational performance (accuracy) of the hybrid pipeline under the modeled network conditions.

The data for training the surrogate grid model are generated in Plexim PLECS \cite{plecs_manual} time-domain simulations. The simulations are performed using fixed-step solver with a step size of 10 \textmu s. Input sequences at DER 1, DER 2, DER 3 and DER 6 consist of uniformly distributed random numbers ranging within [-1, 1] over a duration of 0.1 s and changing every 1 ms. These signals are further scaled by $0.01\mathrm{V_{base}}$ to generate the voltage perturbations injected into the network. The output data are obtained from the integral of current variations measured at DER 8 at a sampling frequency of 1 kHz. To construct the training dataset for the surrogate grid model, the simulation procedure is repeated 1000 times using independently generated input sequences, which is also the data size for surrogate grid model training.

The confusion matrices in Figures \ref{fig_1} and \ref{fig_2} are constructed by counting the number of data samples associated with each combination of predicted labels and ground-truth labels (true labels). The per-class accuracies in Figures \ref{fig_2} and \ref{fig_3} are calculated for each class individually by evaluating the proportion of correctly classified samples among all samples belonging to the corresponding ground-truth class in the original dataset.

\subsection*{Specifications of Multi-Tasking Cases}

Specially, the multi-tasking cases presented in Fig.~\ref{fig_3} are designed to classify two images (denoted as \texttt{Data 1} and \texttt{Data 2}) simultaneously within a single inference process. Compared with the default single-task configuration, the following modifications are introduced:

For both multi-tasking cases, \texttt{Data 1} is generated by shuffling the training/testing dataset using random seed 42, whereas \texttt{Data 2} is generated independently by shuffling using random seed 123.

Spatial multi-tasking (Fig.~\ref{fig_3}a). \texttt{Data 1} is encoded into voltage perturbations applied to DER 1 using Kernel 3 and DER 2 using Kernel 4. \texttt{Data 2} is encoded into perturbations applied to DER 3 using Kernel 3 and DER 6 using Kernel 4. This enables concurrent processing of two independent inputs through spatially separated perturbation channels within the same electric network.

Temporal multi-tasking (Fig.~\ref{fig_3}b). Both \texttt{Data 1} and \texttt{Data 2} are individually encoded into $10\times1$ vectors. The two vectors are concatenated into a single $20\times1$ sequence, where each element is subsequently expanded by repetition over 5 samples rather than 10. This enables sequential multiplexing within the same temporal inference window for the electric network.

For both cases, the output sequence is partitioned into 20 segments instead of 10, with each segment containing 5 elements. The resulting $20\times1$ feature vector is eventually mapped onto 20 classification bins, where the first 10 bins correspond to the prediction of \texttt{Data 1} and the remaining 10 bins correspond to the prediction of \texttt{Data 2}.

The encoders/decoders for both cases are trained using Adam optimizer with learning rate 5\texttimes10\textsuperscript{-3}, batch size 16 and random seed 42.

\clearpage

\clearpage
\begin{table}[ht]
    \centering
    \caption{Voltage Conditions at Buses with DERs}
    \label{tab_DER}
    \begin{tabular}{|p{1cm}|p{2cm}|p{2cm}|p{2cm}|}
        \hline
        \textbf{DER} & \textbf{To Bus} & \textbf{$\mathrm{V}$ [p.u.]}  & \textbf{$\mathrm{\theta}$ [deg]} \\ \hline
        1 & 1   & 1.060 & 0.0 \\ \hline
        2 & 2   & 1.045 & -4.98 \\ \hline
        3 & 3   & 1.010 & -12.72 \\ \hline
        6 & 6   & 1.070 & -14.22 \\ \hline
        8 & 8   & 1.090 & -13.36 \\
        \hline
    \end{tabular}
\end{table}

\begin{table}[ht]
    \centering
    \caption{Load Conditions at Buses}
    \label{tab_load}
    \begin{tabular}{|p{1cm}|p{2cm}|p{2cm}|p{2cm}|}
        \hline
        \textbf{Load} & \textbf{From Bus} & \textbf{$\mathrm{P}$ [MW]}  & \textbf{$\mathrm{Q}$ [MVar]} \\ \hline
        2 & 2   & 21.7  & 12.7 \\ \hline
        3 & 3   & 94.2  & 19.0 \\ \hline
        4 & 4   & 47.8  & -3.9 \\ \hline
        5 & 5   & 7.6   & 1.6 \\ \hline
        6 & 6   & 11.2  & 7.5 \\ \hline
        9 & 9   & 29.5  & 16.6 \\ \hline
        10 & 10  & 9.0   & 5.8 \\ \hline
        11 & 11  & 3.5   & 1.8 \\ \hline
        12 & 12  & 6.1   & 1.6 \\ \hline
        13 & 13  & 13.5  & 5.8 \\ \hline
        14 & 14  & 14.9  & 5.0 \\
        \hline
    \end{tabular}
\end{table}

\begin{table}[ht]
    \centering
    \renewcommand{\thetable}{S3}
    \caption{Line Conditions among Buses}
    \label{tab_line}
    \begin{tabular}{|p{1cm}|p{2cm}|p{2cm}|p{2cm}|p{2cm}|}
        \hline
        \textbf{Line} & \textbf{From Bus} & \textbf{To Bus} & \textbf{$\mathrm{R}$ [p.u.]} & \textbf{$\mathrm{X}$ [p.u.]} \\ \hline
        1 & 1   &   2   & 0.01938   & 0.05917 \\ \hline
        2 & 1   &   5   & 0.05403   & 0.22304 \\ \hline
        3 & 2   &   3   & 0.04699   & 0.19797 \\ \hline
        4 & 2   &   4   & 0.05811   & 0.17632 \\ \hline
        5 & 2   &   5   & 0.05695   & 0.17388 \\ \hline
        6 & 3   &   4   & 0.06701   & 0.17103 \\ \hline
        7 & 4   &   5   & 0.01335   & 0.04211 \\ \hline
        8 & 6   &   11  & 0.09498   & 0.19890 \\ \hline
        9 & 6   &   12  & 0.12291   & 0.25581 \\ \hline
        10 & 6   &   13  & 0.06615   & 0.13027 \\ \hline
        11 & 7   &   8   & 0         & 0.17615 \\ \hline
        12 & 7   &   9   & 0         & 0.11001 \\ \hline
        13 & 9   &   10  & 0.03181   & 0.08450 \\ \hline
        14 & 9   &   14  & 0.12711   & 0.27038 \\ \hline
        15 & 10  &   11  & 0.08205   & 0.19207 \\ \hline
        16 & 12  &   13  & 0.22092   & 0.19988 \\ \hline
        17 & 13  &   14  & 0.17903   & 0.34802 \\
        \hline
    \end{tabular}
\end{table}

\begin{table}[ht]
    \centering
    \caption{Shunt Capacitance Conditions of Lines}
    \label{tab_shunt}
    \begin{tabular}{|p{1.5cm}|p{2cm}|p{2cm}|p{3cm}|}
        \hline
        \textbf{Line} & \textbf{From Bus} & \textbf{To Bus} & \textbf{$\mathrm{B}$ [p.u.]} \\ \hline
        1 & 1   &   2   & 0.0528 \\ \hline
        2 & 1   &   5   & 0.0492 \\ \hline
        3 & 2   &   3   & 0.0438 \\ \hline
        4 & 2   &   4   & 0.0340 \\ \hline
        5 & 2   &   5   & 0.0346 \\ \hline
        6 & 3   &   4   & 0.0128 \\ \hline
        others & --   &   --   & not included \\
        \hline
    \end{tabular}
\end{table}

\begin{table}[ht]
    \centering
    \caption{Transformer Conditions among Buses}
    \label{tab_trafo}
    \begin{tabular}{|p{1cm}|p{2cm}|p{2cm}|p{3cm}|p{2cm}|}
        \hline
        \textbf{Trafo} & \textbf{From Bus} & \textbf{To Bus} & \textbf{Turn Ratio $\mathrm{\alpha}$} & \textbf{$\mathrm{X}$ [p.u.]} \\ \hline
        1 & 4   &   7   & 0.978 & 0.20912 \\ \hline
        2 & 4   &   9   & 0.969 & 0.55618 \\ \hline
        3 & 5   &   6   & 0.932 & 0.25202 \\
        \hline
    \end{tabular}
\end{table}

\begin{table}[ht]
    \centering
    \caption{Label Description of the Datasets}
    \label{tab_dataset}
    \begin{tabular}{|p{1.2cm}|p{3cm}|p{3cm}|}
        \hline
        \textbf{Label} & \textbf{MNIST} & \textbf{MNIST Fashion} \\ \hline
        0 & Handwritten "0"   & T-shirt / Top \\ \hline
        1 & Handwritten "1"   & Trousers \\ \hline
        2 & Handwritten "2"   & Pullover \\ \hline
        3 & Handwritten "3"   & Dress \\ \hline
        4 & Handwritten "4"   & Coat \\ \hline
        5 & Handwritten "5"   & Sandal \\ \hline
        6 & Handwritten "6"   & Shirt \\ \hline
        7 & Handwritten "7"   & Sneaker \\ \hline
        8 & Handwritten "8"   & Bag \\ \hline
        9 & Handwritten "9"   & Ankle Boots \\
        \hline
    \end{tabular}
\end{table}

\begin{table}[ht]
    \centering
    \caption{Dimensions of Trained Parameters in the Study Case}
    \label{tab_par}
    Surrogate Grid Model in comparison with FC Substitution \\
    \vspace{3pt}
    \begin{minipage}[h]{0.45\textwidth}
    \begin{tabular}{|p{2cm}|p{4cm}|}
        \hline
        \multicolumn{2}{|l|}{\textbf{Physics-Informed Surrogate Grid Model}} \\ \hline
        \textbf{Variable} & \textbf{Number of Parameters} \\ \hline
        $\mathbf{u}$ in (\ref{eq_surr_1}) & 8 (incl. \textit{d} and \textit{q}) \\ \hline
        $\mathbf{W}$ in (\ref{eq_surr_2}) & 16 (incl. \textit{dd}, \textit{dq}, \textit{qd}, \textit{qq}) \\ \hline
        $\mathrm{b_{ff}}$ in (\ref{eq_surr_2}) & 2 (incl. \textit{d} and \textit{q}) \\ \hline
        $\mathbf{G}$ in (\ref{eq_surr_3}) & 400 (incl. \textit{dd}, \textit{dq}, \textit{qd}, \textit{qq}) \\ \hline
        $\mathrm{b_{dyn}}$ in (\ref{eq_surr_3}) & 2 (incl. \textit{d} and \textit{q}) \\ \hline
        $\mathbf{H}$ in (\ref{eq_surr_4}) & 400 (incl. \textit{dd}, \textit{dq}, \textit{qd}, \textit{qq})\\ \hline
        $\mathrm{b_{y}}$ in (\ref{eq_surr_4}) & 2 (incl. \textit{d} and \textit{q}) \\ \hline
        $\mathrm{\nu}$ in (\ref{eq_surr_5}) & 1 \\ \hline
        $\mathrm{c}$ in (\ref{eq_surr_5}) & 1 \\ \hline
        $\mathbf{p}$ in (\ref{eq_surr_6}) & 100 \\ \hline
        $\mathrm{b_{z}}$ in (\ref{eq_surr_6}) & 1 \\ \hline
        \textbf{Total} & 933 \\
        \hline
    \end{tabular}
    \end{minipage}
    \hspace{6pt}
    \begin{minipage}[h]{0.45\textwidth}
    \begin{tabular}{|p{2cm}|p{4cm}|}
        \hline
        \multicolumn{2}{|l|}{\textbf{FC Substitution with 30 Hidden Units}} \\ \hline
        \textbf{Module} & \textbf{Number of Parameters} \\ \hline
        400-to-30 & 12,000 weights \\[-1pt]
        FC Layer 1 & 30 bias \\[1pt] \hline
        30-to-10 & 300 weights \\[-1pt]
        FC Layer 2 & 10 bias \\[1pt] \hline
        & \\ \hline
        & \\ \hline
        & \\ \hline
        & \\ \hline
        & \\ \hline
        & \\ \hline
        & \\ \hline
        \textbf{Total} & 12,340 \\
        \hline
    \end{tabular}
    \end{minipage}
    \\
    \vspace{6pt}
    Encoder and Decoder\\
    \vspace{3pt}
    \begin{tabular}{|p{6cm}|p{6cm}|}
        \hline
        \textbf{Module} & \textbf{Number of Parameters} \\ \hline
        FIR Kernel 1 (Conv2d) & 10 (9 weights + 1 bias) \\ \hline
        FIR Kernel 2 (Conv2d) & 10 (9 weights + 1 bias) \\ \hline
        FIR Kernel 3 (Conv2d) & 10 (9 weights + 1 bias) \\ \hline
        FIR Kernel 4 (Conv2d) & 26 (25 weights + 1 bias) \\ \hline
        Linear projection 1 & 6770 (6760 weights + 10 bias) \\ \hline
        Linear projection 2 & 6770 (6760 weights + 10 bias) \\ \hline
        Linear projection 3 & 6770 (6760 weights + 10 bias) \\ \hline
        Linear projection 4 & 5770 (5760 weights + 10 bias) \\ \hline
        Linear Decoder & 20 (10 weights + 10 bias) \\ \hline
        \textbf{Total} & 26,156 \\
        \hline
    \end{tabular}
\end{table}



\clearpage

\section*{Supplementary Note S1: Evaluation on Generalization for More Complex Datasets}

\subsection*{Conditions of the Evaluation:}

The evaluations are performed based on multiple datasets, including the aforementioned MNIST/Fashion-MNIST datasets, and the CIFAR-10 dataset \cite{cifar10}, plus the Butterflies-and-Moths dataset \cite{butterfly} as supplementary results:
\begin{enumerate}
    \item The CIFAR-10 dataset consists of $(32\times32)$ images, 50,000 in the training dataset and 10,000 in the testing one. The images belong to 10 completely mutually exclusive classes.
    
    \item The Butterflies-and-Moths dataset consists of 13,094 training images and 500 test images evenly distributed across 100 different species. In our work, the original $(224\times224)$ images are compressed to $(64\times 64)$ through spatial down-sampling with the \texttt{Pillow} image-processing library in Python.
\end{enumerate}

The \textit{direct classifier} and the \textit{static pass} in Fig.~\ref{fig_5_new} are configured as:

\begin{enumerate}
    \item The \textit{direct classifier} in Fig.~\ref{fig_5_new}(a) is constructed as a 1-dimensional learned convolution which transforms 16 input channels into 10 output channels. Adaptive average pooling is then applied to transform each output feature map into a single scalar value ultimately producing a $1\times10$ vector of class logits. The number of learnable parameters required by the direct classifier is independent of the output size of the last convolutional layer of the feature extractor as well as the input image size.
    
    \item The \textit{static pass} described in Fig.~\ref{fig_5_new}(b) uses the exact same feature extraction and encoding / decoding structure as the \textit{physical computing} pipeline with the physical substrate. However, instead of passing the constructed signal to the four DER input nodes, the static pass instead simply learns a linear combination of the four input signals:
    \begin{align*}
        \mathbf{s}_{\text{out}} = \mathbf{w}^{\top}\mathbf{s}_{\text{in}}+\mathrm{b}, \quad \mathbf{s}_{\text{in}}=\begin{bmatrix}
            \mathbf{s}_1^{\top}\\
            \mathbf{s}_2^{\top}\\
            \mathbf{s}_3^{\top}\\
            \mathbf{s}_4^{\top}
        \end{bmatrix}\in\mathbb{R}^{4\times100}, \ \mathbf{w}\in\mathbb{R}^4, \ \mathrm{b}\in\mathbb{R}^1, \ \mathbf{s}_\mathrm{i}\in\mathbb{R}^{100} \text{ for } \mathrm{i}\in\{1,2,3,4\},
    \end{align*}
    yielding $\mathbf{s}_{\text{out}}\in\mathbb{R}^{100}$ with exactly the same dimensions as the read-out signal produced by the substrate.

    \item Across different feature extractors, the encoding logic remains the same: All feature extractors eventually output 16 feature maps at the last layer, of which the linear combinations result in four flattened 1-dimensional $1\times10$ encoding outputs using adaptive average pooling. The $\mathbf{s}_\mathrm{in}$ in Fig.~\ref{fig_5_new}(b) and $\mathrm{\Delta V}_\mathrm{ref, 1-4}$ in Fig.~\ref{fig_5_new}(c) are subsequently constructed. This encoding logic requires only 68 learnable parameters.
\end{enumerate}

\noindent \textbf{Remark S1-1}: The comparison between the \textit{direct classifier} and the \textit{physical-operator pipeline} assesses whether the hybrid pipeline can use the extracted features competitively under the tested architectures. The comparison between the static-pass and physical-operator pipelines assesses whether the fitted electrical-network transformation improves classification. This comparison does not uniquely attribute any performance difference to a particular nonlinear component, because the feature extractor, encoder, surrogate model and readout procedure can each contribute to the overall mapping.

Among all the tests, two feature extractors are demonstrated:
\begin{enumerate}
    \item Feature extractor designed for grayscale MNIST/Fashion-MNIST datasets, consisting of 2 learnable convolutional layers and 2 ReLU activations, as detailed in Table~\ref{tab:feature_extractor_mnist}.

    \item Improved deeper feature extractor primarily for CIFAR-10 dataset inspired by \cite{silva2023}, consisting of more convolutional layers, ReLU activations and maximum pooling, as detailed in Table~\ref{tab:feature_extractor}.
\end{enumerate}

\subsection*{Evaluation with Butterflies-and-Moths Dataset:}

In this case, model performance is evaluated as an average over 25 independent iterations, where a new model is trained on 10 randomly selected species in each iteration. Similarly, as the complexity of datasets further increase, deeper feature extractors or encoders could contribute to maximizing the physical computationability, which is a straightforward approach towards adapting physical systems for multiple computing tasks.

\clearpage

\begin{figure}[ht]
    \centering
    \includegraphics[width=\linewidth]{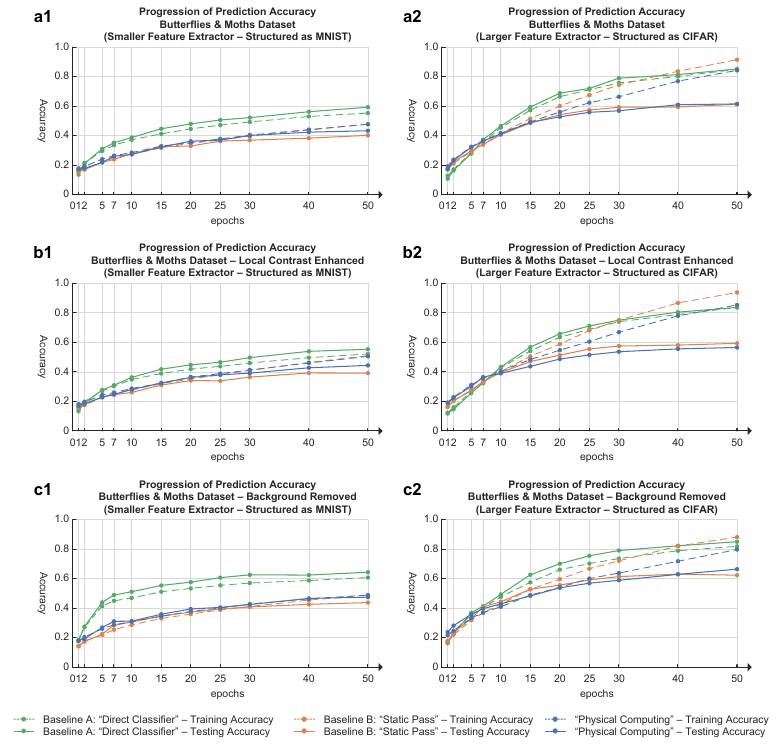}
    \caption{Accuracy progression of models on Butterflies-and-Moths dataset: \textbf{a}. without pre-processing, \textbf{b}. with local contrast enhanced, \textbf{c}. with background removed. Cases are presented where the feature extractor is structured as the MNIST/Fashion-MNIST cases or the CIFAR cases.}
    \label{fig_new_results_butterfly}
\end{figure}

\clearpage

\begin{table}[ht] 
    \centering
    \small
    \renewcommand{\arraystretch}{1.2}
    \caption{Architecture of the direct convolutional classifier in Fig.~\ref{fig_5_new}(a) for $(\mathrm{H\times W})$ images.}
    \label{tab:direct_classifier}
    \begin{tabular}{lcccccc}
        \hline
        \textbf{Layer} & \textbf{Input Ch.} & \textbf{Output Ch.} & \textbf{Kernel} & \textbf{Padding} & \textbf{Activation} & \textbf{Output Size} \\ 
        \hline
        Convolution & 16 & 10 & $1\times1$ & 0 & -- & $\mathrm{10\times H \times W}$ \\
        Adaptive Avg Pooling & 10 & 10 & -- & -- & -- & $\mathrm{10\times 1 \times 1}$ \\ \hline
    \end{tabular}    
\end{table}

\begin{table}[ht]
    \centering
    \small
    \renewcommand{\arraystretch}{1.2}
    \caption{Architecture of the feature extractor for grayscale MNIST input images.}
    \label{tab:feature_extractor_mnist}
    \begin{tabular}{lccccccc}
        \hline
        \textbf{Layer} & \textbf{In Ch.} & \textbf{Out Ch.} & \textbf{Kernel} & \textbf{Padding} & \textbf{Activation} & \textbf{Output Size} & \textbf{Trainable Params.} \\
        \hline
        Convolution 1 & 1 & 32 & $3\times3$ & 1 & ReLU & $32\times28\times28$ & 320 \\
        Convolution 2 & 32 & 16 & $3\times3$ & 1 & ReLU & $16\times28\times28$ & 4,624 \\
        \hline \textbf{Total} & & & & & & & \textbf{4,944} \\
        \hline
    \end{tabular}
\end{table}

\begin{table}[ht] 
    \centering
    \small
    \renewcommand{\arraystretch}{1.2}
    \caption{Architecture of the deeper convolutional feature extractor for $(\mathrm{H\times W})$ images.}
    \label{tab:feature_extractor}
    \begin{tabular}{lccccccc}
        \hline \textbf{Layer} & \textbf{In Ch.} & \textbf{Out Ch.} & \textbf{Kernel} & \textbf{Padding} & \textbf{Activation} & \textbf{Output Size} & \textbf{Trainable Params.} \\ \hline
        Convolution 1 & 3 & 16 & $3\times3$ & same & ReLU & $16\times \mathrm{H}\times \mathrm{W}$ & 448 \\
        Convolution 2 & 16 & 16 & $3\times3$ & same & ReLU & $16\times \mathrm{H}\times \mathrm{W}$ & 2,320 \\
        Max Pooling 1 & 16 & 16 & $2\times2$ & 0 & -- & $16\times \frac{\mathrm{H}}{2}\times\frac{\mathrm{W}}{2}$ & 0 \\ \\
        Convolution 3 & 16 & 16 & $3\times3$ & same & ReLU & $16\times \frac{\mathrm{H}}{2}\times\frac{\mathrm{W}}{2}$ & 2,320 \\
        Convolution 4 & 16 & 16 & $3\times3$ & same & ReLU & $16\times \frac{\mathrm{H}}{2}\times\frac{\mathrm{W}}{2}$ & 2,320 \\
        Max Pooling 2 & 16 & 16 & $2\times2$ & 0 & -- & $16\times \frac{\mathrm{H}}{4}\times\frac{\mathrm{W}}{4}$ & 0 \\ \\
        Convolution 5 & 16 & 16 & $3\times3$ & same & ReLU & $16\times \frac{\mathrm{H}}{4}\times\frac{\mathrm{W}}{4}$ & 2,320 \\
        Convolution 6 & 16 & 16 & $3\times3$ & same & ReLU & $16\times \frac{\mathrm{H}}{4}\times\frac{\mathrm{W}}{4}$ & 2,320 \\ \hline
        \textbf{Total} & & & & & & & \textbf{11,728} \\ \hline
    \end{tabular} 
\end{table}

\begin{table}[ht]
    \centering
    \small
    \renewcommand{\arraystretch}{1.2}
    \caption{Overview of number of trainable parameters per pipeline element.}
    \label{fig:trainable_params_overview}
    \begin{tabular}{lc}
        \hline
        \textbf{Classification Pipeline Element} & \textbf{Num of trainable params.} \\
        \hline
        MNIST feature extractor for grayscale images, 1 in channel & 5k \\
        MNIST feature extractor for RGB images, 3 in channels & 5.5k \\
        CIFAR feature extractor for RGB images, 3 in channels & 12k \\
        \hline
        Encoder & 68 \\
        Decoder & 0 \\
        \hline
        Direct classifier & 170 \\
        \hline
    \end{tabular}
\end{table}

\clearpage
\section*{Supplementary Note S2: Incremental Network-Power Changes Induced by Encoded Perturbations}

\begin{figure}[ht]
    \centering
    \includegraphics[width=0.7\linewidth]{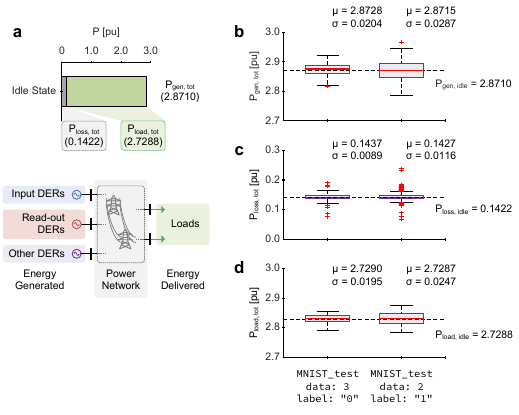}
    \caption{Decomposition of energy footprints through exemplary test cases. Subscripts "\texttt{gen}", "\texttt{loss}", "\texttt{load}" denote the energy/power generated by the DERs, dissipated during transmission and consumed by the loads, respectively. In \textbf{b}-\textbf{d}, power resulting from data encoding is calculated over every 100-$\mu$s windows, with the mean values ($\mu$) and standard deviations ($\sigma$) of the power samples labeled as well.}
    \label{fig_energy_2}
\end{figure}

We quantify the change in modeled network generation, load delivery and transmission loss induced by the voltage perturbations used for inference. This analysis does not provide an end-to-end energy-per-inference comparison with digital hardware, because it excludes the energy required for signal encoding, converter control, sensing, communication and decoding. Instead, it isolates the incremental effect of the encoded perturbations on the modeled power-delivery network in Fig.~\ref{fig_energy_2} relative to its idle operating state.

Seen from the proportion of energy delivered to the loads as desired in the total generated energy, the exemplary MNIST test data lead to 94.99\% (Data 3 with label "\texttt{0}") and 95.03\% (Data 2 with label "\texttt{1}") energy efficiency, respectively, while they should be benchmarked with the idle state energy efficiency 95.05\% instead of the "computing-specific" energy efficiency in digital processors: The energy losses inherently exist in the power transmission networks, and only its variation should be attributed to the computing tasks.

For the two illustrated MNIST samples, the mean additional generated power and transmission losses are of order 
10\textsuperscript{-3}~pu relative to the nominal operating state. These variations are small relative to the baseline power flow in the benchmark model. The result indicates that the specific encoded perturbations do not substantially alter modeled network-level power delivery under the stated assumptions; it does not establish an end-to-end energy advantage over digital inference.

\clearpage
\section*{Supplementary Note S3: KCL Superposition as a Zero-Cost Physical Residual Primitive}

\begin{figure}[ht]
    \centering
    \includegraphics[width=0.7\linewidth]{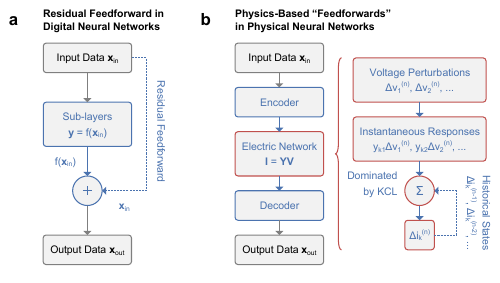}
    \caption{\textbf{Residual superposition in digital neural networks and its inherent physical presence in electric grids.} \textbf{a}, Residual feedforward operation in digital neural networks, where the transformed features $\mathrm{f}(\mathbf{x}_\mathrm{in})$ are superimposed with the input data $\mathbf{x}_\mathrm{in}$, namely $\mathbf{x}_\mathrm{out} = \mathrm{f}(\mathbf{x}_\mathrm{in}) + \mathbf{x}_\mathrm{in}$. \textbf{b}, In a physical neural network as studied, the input voltage perturbations propagate through the electric network, generating instantaneous current responses governed by the Kirchhoff's current law (KCL) as the spatial residual aggregation. Meanwhile, the accumulation of historical states $\mathrm{\Delta i}_\mathrm{k}^\mathrm{(n-1)}$, $\mathrm{\Delta i}_\mathrm{k}^\mathrm{(n-2)}$, ... introduces temporal superposition as a consequence of the memory effects of the electric network.
    }
    \label{fig_4}
\end{figure}

In digital neural networks, residual connections are an explicit architectural choice introduced to address the vanishing gradient problem in deep networks \cite{he2016deep}. Fig.~\ref{fig_4}a shows the digital residual pathway: input data $\mathbf{x}_\mathrm{in}$ branches at the encoder, one path passes through the sublayer transformation $\mathrm{f}(\mathbf{x}_\mathrm{in})$, and the two paths are recombined at an explicit addition node before the decoder. As shown in Fig.~\ref{fig_4}a, the residual operation $\mathbf{x}_\mathrm{out} = \mathrm{f}(\mathbf{x}_\mathrm{in}) + \mathbf{x}_\mathrm{in}$ requires dedicated hardware operations: the input $\mathbf{x}_\mathrm{in}$ must be stored in memory while the sublayer computation $\mathrm{f}(\mathbf{x}_\mathrm{in})$ proceeds, retrieved after the sublayer completes, and then added to the sublayer output through an explicit addition instruction. This sequence — store, compute, retrieve, add — consumes memory bandwidth and clock cycles proportional to the dimensionality of $\mathbf{x}_\mathrm{in}$ at every layer where residual connections appear \cite{rajbhandari2020zero,chen2016training}. 

Fig.~\ref{fig_4}b demonstrates that electric power grids implement an equivalent operation as a direct physical consequence of KCL at zero hardware cost and zero learned parameters. 
Input data $\mathbf{x}_\mathrm{in}$ is encoded as voltage perturbations $\mathrm{\Delta v_{1}^{(n)}}$, $\mathrm{\Delta v_2^{(n)}}$, ... applied simultaneously to multiple input DER nodes. Each perturbation propagates through the admittance network $\mathbf{I = YV}$, generating instantaneous current responses $\mathrm{y_{k1}\cdot\Delta v_1^{(n)}}$, $\mathrm{y_{k2}\cdot\Delta v_2^{(n)}}$, ... at the reading-out bus, where $\mathrm{y_{ki}}$ are the corresponding admittance matrix elements. KCL then enforces the summation of all these contributions at the read-out node — the "$\sum$" node in Fig.~\ref{fig_4}b — as a physical law rather than an arithmetic instruction. The output current $\mathrm{\Delta i_k^{(n)}}$ is therefore the weighted sum of all simultaneous input contributions, implementing the residual aggregation $\mathbf{x}_\mathrm{out} = \mathrm{f}(\mathbf{x}_\mathrm{in}) + \mathbf{x}_\mathrm{in}$ without memory storage, retrieval, or additional hardware.

Fig.~\ref{fig_4} further reveals a second dimension of physical residual aggregation that extends beyond the spatial superposition of simultaneous inputs. The dashed blue arrows entering the KCL summation node from the right represent historical states $\mathrm{\Delta i_k^{(n-1)}}$, $\mathrm{\Delta i_k^{(n-2)}}$, ... — the accumulated current responses from previous voltage perturbation time steps. Since KCL applies equally to all current contributions regardless of when they were generated, the read-out node naturally aggregates both the current instantaneous response and the residual influence of prior inputs. This temporal accumulation emerges from the continuous physical operation of the network across the time-stamped encoding window illustrated in Fig.~\ref{fig_1}c. The result is a physical implementation of both spatial residual aggregation (simultaneous inputs summed by KCL) and temporal residual memory (historical current states accumulated at the reading-out node), that emerges from conservation laws applied to an electrical network.

This dual residual structure — spatial and temporal — is the physical foundation of the multi-task inference capability demonstrated in the following section. Spatial superposition allows multiple independently encoded inputs to coexist in the same inference pass without any interference, since KCL preserves the linear separability of their current contributions at the read-out node. Temporal accumulation allows the network to aggregate information across sequential encoding time steps, enabling the time-division multiplexed inference demonstrated in temporal multi-tasking mode. Both capabilities emerge from the same physical primitive: the conservation of current at every network node, enforced instantaneously and continuously by KCL without architectural design, memory allocation, or learned parameters.

\end{document}